\documentclass[preprint,aps,prl,amssymb,rotating,epsfig,graphics,eepic,showpacs]{revtex4}
\usepackage{epsfig,psfig,rotating,color}
\usepackage{eepic,graphics,graphicx}
\newcommand{\be}{\begin{equation}} \newcommand{\ee}{\end{equation}}
\newcommand{\bea}{\begin{eqnarray}} \newcommand{\eea}{\end{eqnarray}}

\begin{document}


\title{Elastic properties of grafted microtubules}

\author{Francesco Pampaloni$^{1,5}$, Gianluca Lattanzi$^{2,3,4}$,
Alexandr Jon\'a\v{s}$^{5}$, Thomas Surrey$^{1}$, Erwin Frey$^{6}$
and Ernst-Ludwig Florin$^{5}$\footnote{Corresponding author. FP
and GL have equally contributed to this work.}}

\affiliation{$^1$EMBL, Cell Biology and Biophysics Programme,
Meyerhofstr. 1, D-69117 Heidelberg, Germany \\
$^2$Department of Medical Biochemistry, Biology and Physics, Bari, Italy \\
$^3$TIRES-Center of Innovative Technologies for Signal Detection
and Processing, Universit\`a di Bari, Bari, Italy \\
$^4$Istituto Nazionale di Fisica Nucleare, Sezione di Bari, Bari, Italy\\
$^5$Center for Nonlinear Dynamics, University of Texas at Austin,
Austin,
Texas 78712 (USA) \\
$^6$Arnold Sommerfeld Center and CeNS, Department of Physics,
Ludwig-Maximilians-Universit\"at M\"unchen, Theresienstrasse 37,
D-80333 M\"unchen, Germany}

\date{\today}

\begin{abstract}

We use single-particle tracking to study the elastic properties of
single microtubules grafted to a substrate. Thermal fluctuations of
the free microtubule's end are recorded, in order to measure
position distribution functions from which we calculate the
persistence length of microtubules with contour lengths between 2.6
and 48 $\mu$m. We find the persistence length to vary by more than a
factor of 20 over the total range of contour lengths. Our results
support the hypothesis that shearing between protofilaments
contributes significantly to the mechanics of microtubules.  \\

\preprint{LMU-ASC 26/05}

\end{abstract}


\maketitle

\begin{figure} \epsfig{figure=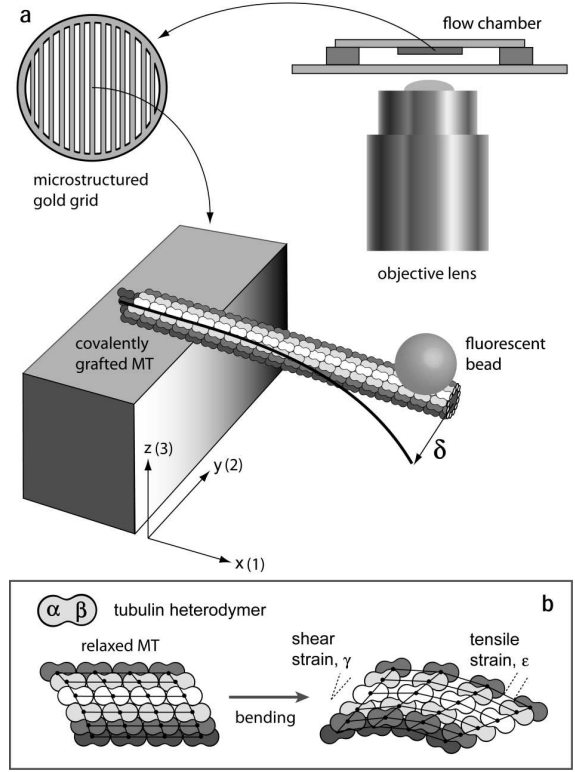,width=7.5cm,clip=}
\caption{\label{fig:exp}Representation of the experimental
situation. (a) The MT is covalently grafted on a gold substrate. A
fluorescent colloidal bead (not to scale) is attached at the MT's
tip, which freely fluctuates in the medium ($\delta \equiv$
deflection). Both $\delta$ and the MT bending profile are
exaggerated for illustration purposes. The distance of the MT from
both sides of the chamber is at least $20$ $\mu$m. (b) Bending
generates a tensile/compressive stress $P=E_{1} \epsilon$ and a
shear stress $s=G_{12} \gamma $ between adjacent protofilaments
($E_1$, $G_{12}$ longitudinal Young's and shear modulus,
respectively; $\epsilon$ tensile strain, $\gamma$ shear strain) .
Given the anisotropy in the tubulin interaction across and along
protofilaments, $E_{1}$ and $G_{12}$ may differ by several orders of
magnitude ($x,y,z$ geometrical axes; $1,2,3$ material axes).}
\end{figure}

\begin{figure}
\epsfig{figure=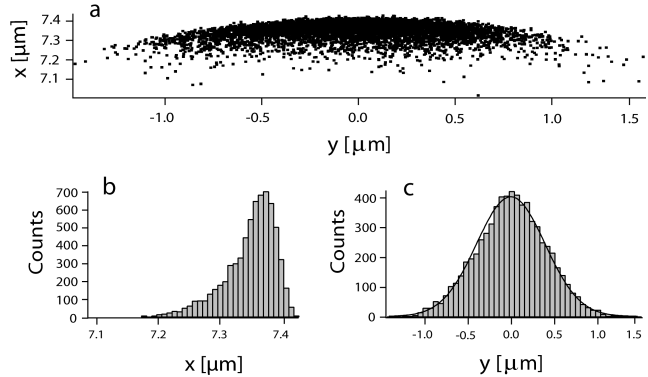,width=8.5cm,clip=} \caption{\label{fig:rdf}
(a) Bead's position tracked over \textbf{$10^4$} frames. The bead
has a diameter of $200$ nm and is attached at the tip of a MT with a
contour length of $7.2$ $\mu$m. (b) Reduced distribution function
along the x-axis (see Fig.~\ref{fig:exp}a). (c) Reduced distribution
function along the y-axis.}\end{figure}

\begin{figure}
\epsfig{figure=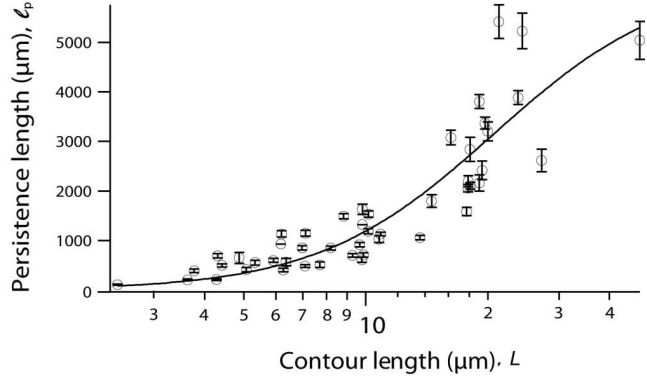,width=8.5cm,clip=} \caption{\label{fig:res}
The persistence length of MTs as a
   function of their contour lengths. The contour length ranges from
   $2.6$ $\mu$m to $48$ $\mu$m. The fit with Eq.~\ref{eq:lp} is
superimposed on the experimental points.}
\end{figure}

The mechanics of living cells is determined by the cytoskeleton, a
self-organizing and highly dynamic network of filamentous proteins
of different lengths and stiffnesses ~\cite{ned03}. Understanding
the elastic response of purified cytoskeletal filaments is
fundamental for the elucidation of the rheological behaviour of the
cytoskeleton. Microtubules (MTs) are hollow cylindrical cytoskeletal
filaments formed by in average thirteen tubulin protofilaments
assembled in parallel. The outer and inner diameters are about $25$
and $15$ nm, respectively. In cells, MTs are generally $1-10$ $\mu$m
long while in axons their length can be $50-100$
$\mu$m~\cite{bra04}. The tubular structure of MTs implies a minimal
cross--sectional area, hence a high strength and stiffness combined
with low density. Although usually modelled as isotropic cylinders,
MTs have anisotropic elastic properties determined by their discrete
protofilament structure. Their response to longitudinal
tensile/compressive stress is determined by the strength of the
head-tail $\alpha \beta $-$\alpha \beta$ tubulin bonds along the
protofilament, whereas the response to shear stress by the weaker
interprotofilament bonds~\cite{van02,dep03,sep03,sch04} which allow
adjacent protofilaments to skew past each other
(Fig~\ref{fig:exp}b). In the language of continuum mechanics and
linear elasticity theory, it is thus expected that the shear modulus
$\emph{G}$ of MTs differs significantly from the Young's modulus
$\emph{E}$~\cite{kas04,kis02}. For instance, mechanically probing
MTs with AFM yields a shear modulus that is 2-3 orders of magnitude
smaller than the Young's modulus ~\cite{kas04, kis02}. Moreover, a
transversal buckling stress of $P_r = 600$ Pa has been recently
measured on MTs~\cite{nee04}. This is four orders of magnitude
smaller than the buckling stress calculated by modelling MTs as
isotropic solids with $E \sim 1.6$ GPa. The main parameter for the
description of the mechanics of polymer filaments is the ratio $
\ell_p / L$ between the persistence length $\ell_p$ and the contour
length $L$, which quantitatively describes the stiffness of a
filament. $\ell_p$ is expected to depend on both $G$ and $E$, but a
$G \ll E$ implies also a dependence of $\ell_p$ on $L$, as suggested
by the recent AFM experiments in~\cite{kis02}. The most sensitive
and elegant way to determine $\ell_p / L$ is measuring the
probability distribution function of the end-to-end distance of the
filament, $P(R)$~\cite{wil96}. $P(R)$ and $ \ell_p / L$ have been
measured with great accuracy for actin~\cite{leg02}. However, a
detailed study of the mechanics of MTs based on probability
distributions is complicated by their high stiffness, requiring an
assay with very high spatial resolution. In this Letter, we describe
a method based on single-particle tracking (SPT) for measuring the
probability distribution function $P(x,y)$ of grafted MTs. From the
measured probability distributions we obtain the MTs persistence
length and investigate its dependence on $L$ over a broad range of
contour lengths ($2.6 - 48 \mu$m). In the assay, single fluorescent
colloidal beads are attached at the tip of grafted MTs and the
thermal fluctuations of the MT-bead system are recorded. Indeed,
thermal fluctuations are the most natural way to mechanically probe
biological nanometer-sized specimens non-destructively and over
extended times. The experimental situation is depicted in
Fig~\ref{fig:exp}a. Biotinylated, rhodamine-labelled MTs,
polymerized according to a standard protocol~\cite{car92} and
stabilized with taxol are covalently attached to a microstructured
gold substrate made of parallel bars (bar width $10$ $\mu$m,
distance between the bars $50$ $\mu$m) (Fig~\ref{fig:exp}a). The
gold surface is chemically modified to ensure well-defined and
reproducible conditions for the attachment of MTs~\footnote{The grid
surface is cleaned with a solution H$_2$O$_2$/NH$_3$/H$_2$O
$1$:$1$:$5$ and incubated overnight with mercaptoundecanoic acid
($5$ mM in ethanol), producing a self-assembled monolayer with
exposed carboxylic groups. These are activated by an aqueous
solution of N-hydroxysuccinimide and
1-ethyl-3-(3-dimethylaminopropyl)carbodiimide both $100$ mM to
ensure the covalent attachment of MTs.}. A flow cell is assembled as
shown in Fig~\ref{fig:exp}a, setting a distance  of $20-30$ $\mu$m
between the gold substrate and the upper and lower boundaries of the
cell, so that any influence of the chamber's walls is negligible. A
MTs suspension is then flowed into the cell perpendicularly to the
bars, inducing MTs to attach preferentially along this direction,
with just a segment of MT grafted to the grid, while the free
portion fluctuates unconstrained in the medium (Fig~\ref{fig:exp}a).
A diluted suspension of yellow-green fluorescent, avidin-coated
colloidal beads (diameter $200$ or $500$ nm, Molecular Probes F-8774
and Bang Laboratoires CP01F, respectively) is then flowed into the
chamber. The beads bind strongly to the biotinylated MTs. Finally,
the chamber is sealed with silicon grease, to avoid evaporation of
the medium over time. In living cells, MTs usually irradiate from a
microtubule organizing center toward the cell cortex ~\cite{ned03},
hence the boundary conditions adopted in our assay are close to MT
configurations in cells. Single grafted MTs with one bead attached
proximally to the tip are selected for the measurement.
Alternatively, optical tweezers are employed to attach the bead to
the MT's tip. Since the beads and the MTs emit fluorescence at two
different wavelengths, they can be observed independently by
exchanging the filter set in the microscope. Fig.~\ref{fig:rdf}a
shows an example of the distribution resulting from the projection
of the three-dimensional fluctuations of the MT's tip on the $xy$
plane. The distribution is obtained by measuring the position of the
fluorescent bead attached at the MT's end in at least $10^4$ frames
(average frame rate $16$ frames/s). The bead's position is
automatically tracked frame-by-frame with a custom particle-tracking
routine. The coordinates of the bead's centroid can be determined
with an accuracy of $1$ nm and better under optimal conditions
~\cite{spe03}. Here, the accuracy was slightly reduced to $\sim10$
nm. A stack containing the fluorescent MTs is recorded separately
for the determination of $L$.
 The recorded particle
position is used to obtain the probability distribution function
$P(x,y)$ of the free end in the $xy$ plane and the corresponding
reduced distribution functions $P(x)$ and $P(y)$
(Fig.~\ref{fig:rdf}b-c), obtained by integrating $P(x,y)$ over the
coordinates $y$ and $x$ respectively~\footnote{According
to~\cite{leg02} the longest relaxation time for a semiflexible
polymer is $\tau_C = (\zeta /\ell_p k_B T) (L/A)^4$, where $\zeta =
2.47 \times 10^{-3}$ Pa s is an effective friction coefficient and
$A = 1.875 $ for a polymer with one clamped and one free end. We
find that $\tau_c = 76.6$ ms for a MT of 10 $\mu$m: end-to-end
fluctuations saturate quite fast to the equilibrium value within the
time of the observation}. These can be directly compared with
theoretical predictions based on the wormlike chain model, as
reported in~\cite{lat04}. Given the high persistence length of a
microtubule ($\ell_p/L$ can be as high as several hundreds), the
longitudinal distribution function $P(x)$ is peaked towards full
stretching, as in Fig.~\ref{fig:rdf}c, and varies over a typical
length scale $L_\parallel = L^2/\ell_p$. A non-Gaussian and
asymmetric shape of $P(x)$, is clearly visible in
Fig.~\ref{fig:rdf}b. In contrast, $P(y)$ is a Gaussian whose width
is given by the transverse length scale $L_\perp = \sqrt{(2 L^3 / 3
\ell_p)}$, as represented in Fig.~\ref{fig:rdf}c
~\cite{ben03,wil:unpub}. By measuring the contour length $L$,
$\ell_p$ can be extracted from both $P(x)$ or $P(y)$. It is
important to notice that the measured $P(x)$ and $P(y)$ correspond
to convolution products between the theoretical functions
$P_{th}(x)$ and $P_{th}(y)$ predicted by the wormlike chain
model~\cite{ben03,wil:unpub} and the function that represents the
experimental precision in the measurement of the distance between
the tip of the filament and the attachment point on the substrate.
Since longitudinal and transverse fluctuations scale differently,
the convolution has a negligible effect on $P_{th}(y)$, in contrast
to $P_{th}(x)$ whose shape is much more sensitive to small
variations of the experimental precision. For this reason, we
focused on $P(y)$ for the determination of $\ell_p$ in this work.
The persistence lengths obtained from $P(y)$ for $48$ grafted
microtubules of different contour lengths are shown in
Fig.~\ref{fig:res}. We observe a clear dependence of $\ell_p$ on
$L$. This result is a direct consequence of the anysotropic tubulin
interactions along and across protofilaments. In facts, as a
consequence of their anisotropic protofilament structure, MTs are
expected to have five independent elastic moduli across the
longitudinal ($xy$) and transversal ($yz$) planes (i.e. $E_{1}$,
$E_{2}=E_{3}$, $G_{12}=G_{13}$, $G_{23}$, and the Poisson's ratios
$\nu_{13}=\nu_{12}$; $1,2,3$ are the material axes,
Fig.~\ref{fig:exp}a), similarly to fiber-reinforced materials
~\cite{kol03}. In the following discussion we will model MTs as
anisotropic hollow cylindrical beams and assume that the small
fluctuations in the tip's position can be interpreted as deflections
due to thermal forces. Given the boundary conditions and the
geometry of the system, only the elastic moduli $E_{1}$ and $G_{12}$
are involved in bending on the $xy$-plane. Microscopically, $E_{1}$
is expected to arise from the strong $\alpha \beta $-$\alpha \beta$
tubulin bonds along single protofilaments, whereas the origin of
$G_{12}$ is expected to be in the much weaker interaction between
tubulin dimers on adjacent protofilaments (see Fig.~\ref{fig:exp}b),
which are allowed to skew past each other in order to accommodate
the shear strain associated with MTs bending. Thus, the total
deflection of MTs ($\delta$) is the sum of both bending ($\delta_B$)
and shear deformations ($\delta_S$)~\cite{tim51,kis02}:

\be \delta = \delta_B + \delta_S = \frac{PL^3}{\alpha E_{1} I_2} +
\frac{P L}{G_{12}k A } = \frac{PL^3}{\alpha \ell_p k_B T},
\label{eq:elastic} \ee

\noindent where $P$ is the lateral load at the MT's tip, $I_2 =
(1/4) \pi (R_e^4 - R_i^4)$ is the second moment of the tube's
cross-sectional area $A$, $k$ a geometrical correction factor whose
value is $0.72$ for a hollow cylindrical beam~\cite{kis_b03}, and
$\alpha=3$ for a grafted beam. The last equality, with $k_BT$ the
thermal energy, represents the linear response of a semiflexible
polymer~\cite{kro96}. Through the relation $E_{1} I_2 = \ell_p k_B
T$, Eq.~\ref{eq:elastic} can be used to obtain the length dependence
of the persistence length:

\be \ell_p = \ell_p^\infty \left( 1 + \frac{3 E_{1} I_2}{G_{12} k A
L^2} \right)^{-1}, \label{eq:lp} \ee

\noindent where $\ell_p^\infty$ is the persistence length for long
microtubules ($L \gg \sqrt{3 E_{1} I_2/G_{12} k A} \simeq  21
\mu$m).

Eq.~\ref{eq:lp} provides a good fit to the experimental points,
giving a value of $\ell_p^\infty = 6.3  \pm  0.8 $ mm in agreeement
with previous measurements on long MTs~\cite{git93}. This value
corresponds to $E_{1}= 1.51 \pm 0.19 $ GPa and a ratio $\delta_S /
\delta_B = (21/L)^2$ when $L$ is expressed in micrometers,
corresponding to a ratio between Young's and shear modulus of $\sim
10^{6}$, three orders of magnitude higher than previously
reported~\cite{kas04,kis02}. This can be partially explained by the
fact that in ref.~\cite{kas04,kis02} MTs were stabilized with
glutharaldeyde, a cross-linker used for chemical fixation of
proteins, whereas we use the drug taxol, which insert in the MTs
lattice and may relieve the internal tension inducing a further
decrease of the shear modulus~\cite{fel96}. In addition, experiments
reported in ref.~\cite{kas04,kis02} suggest a strong dependence of
the shear modulus on temperature. Whereas $E\sim G$ in conventional
materials, a shear modulus significantly smaller than the Young's
modulus is typical of biological materials (from wood to bones) and
engineered composites. Down to nanometer scale, a ratio $E\setminus
G\sim10^{2}$ has also been measured for single-walled carbon
nanotubes~\cite{sal99}.
 As shown in Fig.~\ref{fig:res}, the values of $\ell_p$ measured on
MTs spread with increasing $L$. This spreading is partially due to
differences in the structures of single MTs. In facts, in MTs
polymerized \emph{in vitro} the number of protofilaments may vary
from one MT to another and even along a given MT, and several
lattice types are possible~\cite{chr92}. The probability of defects
or inhomogeneities in the MT lattice may also increase with the
contour length. Since there are also evidences that the stiffness of
MTs can vary with the polymerization velocity~\cite{jan04}, we
carefully monitored the MT polymerization temperature. A further
potential experimental source of error might be a loose binding of
the bead to the MT, which could lead to an underestimation of
$\ell_p$. This has been tested by pulling the bead with optical
tweezers at high laser power, ensuring that it attaches firmly to
the MT and does not roll or twist on it. In addition, by strongly
bending the MT with optical tweezers we ensured that the segment of
the MT anchored to the substrate does not change its orientation
during thermal fluctuations, a further source of uncertainty in the
measurement of $L$. Theoretical calculations show also that the
fluctuations of the angle of torsion along the whole microtubule are
below 7 degrees and can be safely neglected. In summary, we
developed an assay based on single-particle tracking to measure with
high spatial resolution the reduced distribution functions $P(x)$
and $P(y)$ for the tip's position of grafted MTs. Through a
systematic measure of the persistence length of MTs with contour
lengths between \textbf{$2.6$} and \textbf{$48$} $\mu$m, we observe
a dependence of the MTs stiffness on $L$. For short MTs ($L=2-3\mu
m$) $\ell_p$ is at least one order of magnitude smaller than for
long ones ($L\geq30-40\mu m$). This property can be explained by the
anisotropic structure of MTs, and described by linear elasticity
theory. Our results together with those in~\cite{nee04, kas04,
kis02} suggest that the interprotofilament bonds are much weaker
than the $\alpha \beta $-$\alpha \beta$ tubulin bonds along the
protofilaments, leading to a very low shear modulus in MTs. The
dependence of $\ell_p$ on $L$ in MTs might also contribute to
explain the broad range of measured MTs persistence lengths found in
literature ($0.5 - 5$ mm)~\cite{fel96,git93,jan04}. Noticeably, the
maximum variability of $\ell_p$ occurs at contour lengths close to
the typical length scale of a cell ($L\simeq12$ $\mu$m). It is
interesting to speculate about the biological significance of this
property. MTs are able to accomplish various tasks in differently
sized cells and during different stages of a cell's life cycle. On
one hand, flexibility of MT's tip allows MTs to search space
laterally for binding partners or to bend and continue growing in a
modified direction when encountering obstacles, for example when
growing MTs hit the side of a small fission yeast cell. On the other
hand, sufficient rigidity is important in situations in which MTs
need to resist to pushing forces such as in large vertebrate cells
during anaphase of mitosis, when the spindle elongates. The length
dependence of MTs rigidity might have allowed cells during evolution
to use MTs in an optimal way both for lateral searching and for
exerting pushing forces in different contexts and especially on
different length scales. We are grateful to A. Rohrbach, E. H. K.
Stelzer and J. Swoger for their substantial support and many
interesting discussions. GL's research has been partially supported
by a Marie Curie Fellowship under contract no.~HPMF-CT-2001-01432.
FP has been supported by the German Research Foundation, grant no.
FL351/2-1.


\end{document}